\documentstyle[prl,aps,epsf]{revtex}         
\begin{document}
\draft               
\twocolumn[\hsize\textwidth\columnwidth\hsize\csname @twocolumnfalse\endcsname

\title{Evidence for a critical velocity in a Bose--Einstein condensed
gas}
\author{C. Raman, M. K\"{o}hl, R. Onofrio, D.S. Durfee, C.E. Kuklewicz,
Z. Hadzibabic, and W. Ketterle}
\address{Department of Physics and Research Laboratory of
Electronics, \\
Massachusetts Institute of Technology, Cambridge, MA 02139}
\date{\today{}}
\maketitle
\begin{abstract}
We have studied dissipation in a Bose--Einstein condensed gas by moving
a blue detuned laser beam through the condensate at different
velocities.  Strong heating was observed only above a critical velocity.
\end{abstract}
\pacs{PACS 03.75.Fi, 67.40.Vs,67.57.De}
\vskip1pc
]
Macroscopic quantum coherence and collective excitations are key
features in our understanding of the phenomenon of superfluidity.
The superfluid velocity is proportional to the gradient of the
phase of a macroscopic wavefunction.  Collective excitations 
determine a critical velocity below which the flow is dissipationless.  
This velocity is given by Landau's criterion\cite{landau},
\begin{equation}
v_{c} = \min \left( \frac{\varepsilon (p) }{p} \right )
\end{equation}
where $\varepsilon$ is the energy of an excitation with momentum $p$.
Critical velocities for the breaking of Cooper pairs in $^{3}$He and
the generation of rotons\cite{rotons} and vortices\cite{orifices} in
$^{4}$He have been extensively studied.

Bose-Einstein condensed gases (BEC) are novel quantum
fluids\cite{varenna}.  Previous work has explored some aspects
related to superfluidity such as the macroscopic phase
\cite{interference} and the phonon nature of low-lying collective
excitations\cite{varenna,phonon bragg}.  In this Letter we report
on the measurement of a critical velocity for the excitation of a
trapped Bose-Einstein condensate.  In analogy with the well known
argument by Landau  and the vibrating wire experiments in
superfluid helium \cite{wires}, we study dissipation when an
object is moved through the fluid. Instead of a massive
macroscopic object we used a blue detuned laser beam which repels
atoms from its focus to create a moving boundary condition.

The experiment was conducted in a new apparatus for the production
of Bose-Einstein condensates of sodium atoms.  The cooling
procedure is similar to previous work \cite {cloverleaf}---the new
features have been described elsewhere \cite{onofrio:99}. Briefly,
laser cooled atoms were transferred into a magnetic trap in the
Ioffe-Pritchard configuration and further cooled by rf evaporative
cooling for 20 seconds, resulting in condensates of between 3 and
12 $\times 10^{6}$ atoms. After the condensate was formed, we
reduced the radial trapping frequency to obtain condensates which
were considerably wider than the laser beam used for stirring.
This decompression was not perfectly adiabatic, and heated the
cloud to a final condensate fraction of about 60\%. The final
trapping frequencies were $\nu_{r}=65$ Hz in the radial and
$\nu_{z}=18$ Hz in the axial direction.  The resulting condensate
was cigar-shaped with Thomas-Fermi diameters of 45 and 150 $\mu$m
in the radial and axial directions, respectively.  The final
chemical potential, transition temperature $T_{c}$ and peak
density $n_{0}$ of the condensate were 110 nK, 510 nK and $1.5
\times 10^{14} \text{cm}^{-3}$, respectively.

The laser beam for stirring the condensate had a wavelength of 514 nm
and was focused to a Gaussian $1/e^{2}$ beam diameter of $2w = 13
\mu$m.  The repulsive optical dipole force expelled the atoms from the
region of highest laser intensity.  A laser power of 400 $\mu$W created
a 700 nK barrier resulting in a cylindrical hole $\sim 13 \mu$m in
diameter within the condensate.  The laser barrier created a {\em
soft} boundary, since the Gaussian beam waist was more than 10 times
wider than the healing length $\xi = (8 \pi a n_{0})^{-1/2} = 0.3
\mu$m, $a$ being the two-body scattering length.

\begin{figure}[htbf]
\epsfxsize=70mm
\centerline{\epsfbox{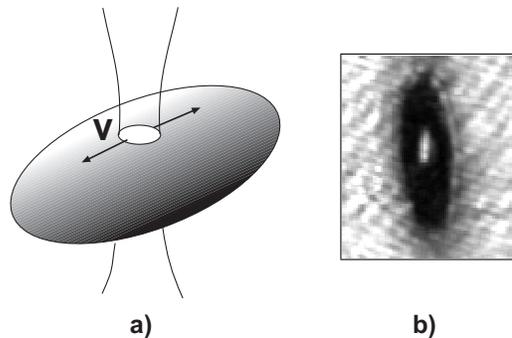}}
\caption{Stirring a condensate with a blue detuned laser beam. a) The laser beam
diameter is $13 \mu$m, while the radial
width of the condensate is $45 \mu$m.  The aspect ratio of the
cloud is $3.3$. b) In situ absorption image of a condensate with
the scanning hole. A 10 kHz scan rate was used for this image to
create the time-averaged outline of the laser trajectory through
the condensate.}
\end{figure}

The laser was focused on the center of the cloud.  Using an
acousto-optic deflector, it was scanned back and forth along the
axial dimension of the condensate (fig. 1).  We ensured a constant
beam velocity by applying a triangular waveform to the deflector.
The beam was scanned over distances up to $60 \mu$m, much less
than the axial extent of the condensate of $150 \mu$m.  Therefore,
axial density variations were small within the scan range.  The
scan frequencies were varied by a factor of 3 between 56 and 167
Hz.  Scan velocities close to the speed of sound required scan
frequencies much larger than the axial trapping frequency
$\nu_{z}$.

After exposing the atoms to the scanning laser for 900 ms, we
allowed the cloud to equilibrate for 100 ms, then turned off the
magnetic trap and recorded the time-of-flight distribution after 35 ms
on a CCD camera using near-resonant absorption imaging.  The
condensate fraction $N_{0}/N$ was determined from fits
to the bimodal velocity distribution\cite{fraction}.  We found that the decreasing
condensate fraction was a more robust measure of the heating induced
by the moving laser beam than the temperature extracted from the wings
of the thermal cloud.  No heating was observed when the laser beam was
kept stationary for the entire 900 ms.

Figure 2 shows the effect of the moving laser on the condensate
for three different scan rates $f = 56, 83$ and $167$ Hz.  The
heating rate is higher (larger final thermal fraction) for larger
drive amplitudes $d$ and higher scan frequencies $f$.  When the
same data are replotted as a function of the {\em velocity} of the
laser beam $ 2 d f $ (fig.  3), two features emerge immediately.
First, all the three data sets collapse onto one universal curve,
indicating that the heating of the condensate depends primarily on
the velocity of the beam, and {\em not} on either frequency or
amplitude independently.  This suggests that the observed
dissipation is not strongly affected by the trapping potential and
discrete resonances, but rather, reflects bulk properties of the
condensate.

\begin{figure}[htbf]
\epsfxsize=70mm
\centerline{\epsfbox{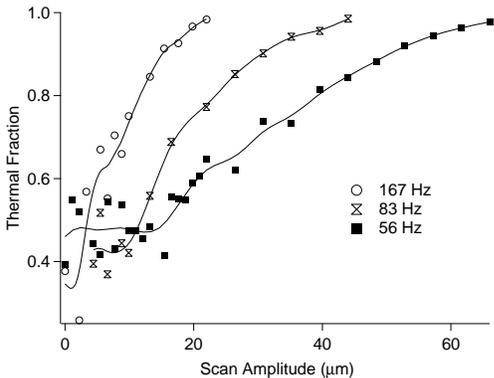}}
\caption{ Heating a condensate with a moving laser beam.  The
final thermal fraction increased with both the scan amplitude and
the scan rate of the laser beam.  The total exposure time was the
same for all data points.  They represent single shot
measurements.  Solid lines are smoothing spline fits to guide the
eye.  }
\end{figure}

The second feature is that we can distinguish two regimes of
heating separated by the dashed line in fig. 3.  For low
velocities, the dissipation rate was low and the condensate
appeared immune to the presence of the scanning laser beam. For
higher velocities, the heating increased, until at a velocity of
about 6 mm/s the condensate was almost completely depleted for a
900 ms exposure time. The cross-over between these two regimes was
quite pronounced and occurred at a velocity of about $1.6 $ mm/s.
This velocity should be compared with the speed of sound in the
condensate. Since a condensate released from the magnetic trap
expands with a velocity proportional to the speed of sound, we
could determine its value directly from time-of-flight absorption
images to be 6.2 mm/s (at the peak density), almost a factor of 4
larger than the observed critical velocity.

\begin{figure}[htbf]
\epsfxsize=70mm
\centerline{\epsfbox{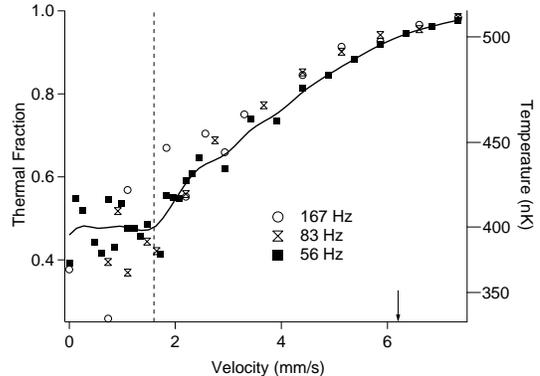}}
\caption{ Evidence for a critical velocity.  The same three data
sets as in figure 2 are replotted vs. laser beam velocity and all
appear to lie upon a universal curve.  The dashed line separates
the regimes of low and high dissipation.  The right axis indicates
the temperature $T= (1-N_{0}/N)^{1/3} T_{c}$.  The peak sound
velocity is marked by an arrow.  The data series for 83 and 167 Hz
showed large shot-to-shot fluctuations at velocities below 2
mm/sec.  The solid line is a smoothing spline fit to the 56 Hz
data set to guide the eye.}
\end{figure}

To rule out the possibility of heating through the non-condensed
fraction, a control experiment was performed on clouds at two
different temperatures above $T_{c}$, 800 nK and 530 nK, the latter
quite close to the transition temperature.  No heating of the cloud
was observed in either case, for scan velocities of up to 14 mm/s;
this reflects the small overlap between the scanning laser and the
large thermal cloud.  Since these clouds are typically not in the
hydrodynamic regime of collisions, we used a single particle model for
heating based on collisions with a moving wall to scale our
measurements from above to below $T_{c}$.  We obtained a conservative
upper bound of 15\% of the observed
temperature rise which could be attributed to the non-condensed fraction.

What determines the critical velocity?  In a quasi-homogeneous
dilute Bose gas near zero temperature, the critical velocity for
phonon excitation is the speed of Bogoliubov sound, which depends
on the density $n(r)$ through
\begin{equation}
c_{B}(r) = \sqrt {\frac{4 \pi \hbar^2 a}{M^2} n(r)}
\end{equation}
where M is the atomic mass.  Alternately, one may expect that the
rapid flow around the laser beam generates vorticity in the fluid.
For the excitation of vortex pairs in a channel of diameter D, eq. (1)
leads to a critical velocity \cite{feynman}

\begin{equation}
v_{c}  = \frac{\hbar }{{M D}}\ln \left( {\frac{D}{\xi }} \right).
\end{equation}

Since this velocity scales inversely with the size of the system, it
is lower than the value predicted by eq. 2.  For our situation eq. 3
can only give an approximate estimate due to the inhomogeneous density
and the different geometry.  For a channel width of $15 \mu$m and
a healing length of $0.3 \mu$m it yields a critical velocity of about 0.7 mm/s,
a factor of 2 below our measurement.

The criterion given in eq. 3 is only based on considerations of energy and
momentum and provides a lower bound to $v_c$.  In addition, one must
consider how to produce the excitations dynamically.  Several
mechanisms for creating vortex lines have been discussed, including
remanent vorticity and vorticity pinned to the surface
\cite{wilks}. In the latter case, the critical velocity depends on the
surface roughness.  The gaseous condensates are confined to magnetic
traps which provide perfectly smooth boundary conditions.  Therefore,
we expect that the critical velocity is determined by nucleation
dynamics rather than by purely energetic arguments.

The relevant criterion for the onset of dissipation in a Bose
condensed gas obeying the nonlinear Schr\"{o}dinger equation has been
discussed in several papers.  According to these theories, dissipation
ensues when the relative velocity between the object and the fluid
exceeds the speed of sound $c_{B}$ {\em locally}
\cite{frisch:92,huepe,adams:99}. For an incompressible flow around a
cylindrical object this velocity peaks at the side, reaching twice the
object's speed. The hydrodynamic equations for the compressible
condensate imply that the faster the velocity of the flow field, the
lower is the condensate density. This effect lowers the critical
velocity for a cylindrical object even further\cite{frisch:92}:
\begin{equation}
v_{c} = \sqrt\frac{2}{11} c_{max} = 0.42 c_{max}
\end{equation}

This result is independent of the size of the object and was
corroborated by numerical simulations of the nonlinear
Schr\"{o}dinger equation in a homogeneous gas
\cite{frisch:92,huepe,adams:99}. For our conditions, this estimate
yields a critical  velocity of $2.6$ mm/s, or $1.6$ times the
observed threshold.  However, the finite size of the condensate in
our experiments\cite{finite}, its inhomogeneous density
distribution and the soft boundary of the laser beam are not
accounted for by the theory. All these effects should lower the
critical velocity.

What happens to the condensate above $v_{c}$?  Numerical
simulations of the nonlinear Schr\"{o}dinger equation were used
to study the flow field around an object moving through a
homogeneous condensate\cite{frisch:92,huepe,adams:98}.  These
studies show that above a critical velocity given by Eq.  (4) the
superfluid flow becomes unstable against the formation of
quantized vortex lines, which signals the onset of a new,
dissipative regime.  Pairs of vortices with opposite circulation
are generated at opposite sides of the object in order to reduce
the high local flow speed.  The rate of heating can be estimated
from the energy of vortices and the vortex shedding frequency.

The energy of a vortex pair $\varepsilon_{vp}$ is estimated with
the assumption that the vortices are separated by the object
diameter $2w$\cite{thouless},

\begin{equation} \varepsilon_{vp} = 2 \pi D  \frac{ n\hbar^{2}}{M}
\ln
\frac{2 w}{\xi}
\end{equation}
where $D$ is the radial width of the condensate and $n$ is the
density. This yields a value of about 3.4 mK or 1.3 nK/atom.
Numerical simulations \cite{frisch:92,adams:99} have shown that
for $v>v_{c}$ the rate of vortex pair shedding is proportional to
$v-v_{c}$.  The proportionality constant, besides a numerical
factor, is the mean field energy in frequency units divided by the
speed of sound.  Thus the rate of change in temperature should
have the form $\dot{T} = \kappa (v-v_{c})$. Using the estimate for
the vortex energy, the model predicts $\kappa \approx 160$ nK/mm,
in rough agreement with our measured heating rate near the
threshold which gives $\kappa = 62$ nK/mm.

In conclusion, we have established a method for studying
dissipation in a Bose condensate by implementing a scanning
``hole'' induced by a far-off-resonant laser beam.  Both the laser
beam and the trapped condensate provide clean boundary conditions.
This and the simplicity of the system makes it amenable to
theoretical treatments of vortex nucleation and dissipative
dynamics.  We found evidence for a critical velocity for
excitation of the condensate.  Both the onset and the magnitude of
the observed heating are in qualitative agreement with model
calculations based on the non-linear Schr\"{o}dinger equation
which predict dissipation when the flow field becomes locally
supersonic. In contrast, a similar study on rotational stirring
concluded that the speed of sound near the stirrer is irrelevant
for vortex nucleation \cite{burnett}. In further studies, we plan
to vary the geometry and density in order to distinguish between
different predictions for critical velocities which depend only on
geometry (eq. 3), or only on the density (eq. 4). All the
calculations were done in two dimensions and at zero temperature.
In the experiment, the laser beam passes through the surface of
the condensate where the density vanishes.  Because of this and
the non-zero temperature, we expect finite dissipation even at low
velocities and a smooth crossover between low and high
dissipation.  More precise measurements of the heating should
allow us to study these finite-size and finite-temperature
effects.

We thank A. Chikkatur and A. G\"{o}rlitz for experimental assistance
and L. Pitaevskii, J.C. Davis and G. Pickett for useful discussions on
critical velocities in liquid helium.  This work was supported by the
ONR, NSF, ARO, NASA, and the David and Lucile Packard Foundation.
M.K. also acknowledges support from Studienstiftung des Deutschen
Volkes.


\begin{thebibliography}{99}
\bibitem{fraction}
    Data were excluded when the total number of atoms N deviated by
    more than +/- 25 \% of the average.
\bibitem{finite}
   The radial extent of the condensate $D = 45 \mu$m is only 3.5
    times the laser beam diameter.  This should  constrict the flow
    around the laser beam leading to faster flow velocities.
\end{thebibliography}
\end{document}